\documentclass{elsart}

\usepackage{natbib}

\usepackage{amsmath}

\begin{document}

\begin{frontmatter}


%

\title{Chiral Symmetry in Dirac Equation and its Effects on
Neutrino Masses and Dark Matter}

\author{T.B. Watson and Z. E. Musielak}
\address{Department of Physics, The University of Texas at 
Arlington, Arlington, TX 76019, USA}
%

\begin{abstract}
Chiral symmetry is included into the Dirac equation using the irreducible 
representations of the Poincar\'e group.  The symmetry introduces the 
chiral angle that specifies the chiral basis.  It is shown that the correct 
identification of these basis allows explaining small masses of neutrinos 
and predicting a new candidate for Dark Matter massive particle. 
\end{abstract}

\end{frontmatter}

\section{Introduction}

The work of Wigner [1] was among the first to highlight the crucial role of group 
theory in Quantum Field Theory (QFT).  He classified the irreducible representations 
(irreps) of the Poincar\'e group and identified an elementary particle with an unitary 
irrep of the group [2];  the explication of single particle states as the irreps of the 
Poincar\'{e} group has done much to validate the physical foundations of modern 
particle physics (e.g., [3]).  As elucidated in previous works [4-7], such group 
theoretical notions may be utilized to define a fundamental theory as satisfying the 
following  principles: the principle of Poincar\'e and gauge invariance, the principle 
of locality, and the principle of least action.

Among the fundamental equations of QFT, the Dirac equation [8] plays a special 
role because it describes fermionic fields, which include quarks and leptons of the 
Standard Model of particle physics [3].  The original Dirac method to obtain this 
equation [9] is often replaced in QFT textbooks by giving the required Lagrangian 
but without any explanation of its first principle origin (e.g., [10]).  Other methods 
to derive the Dirac equation include the Lorentz transformations of four-component 
spinors [3] and the Bargmann-Wigner approach [4] that is based on the Poincar\'e 
group of the Minkowski metric of QFT.  The general structure of the Poincar\'{e} 
group is $\mathcal {P}\ =\ SO (3,1) \otimes_s T (3+1)$, where $SO(3,1)$ is a 
non-invariant Lorentz group of rotations and boosts and $T(3+1)$ an invariant 
subgroup of spacetime translations, and this structure includes reversal of parity 
and time [2].

In this paper, we present a novel method that uses the irreps of $T(3+1)$ to derive 
the Dirac equation with chiral symmetry.  There are two main aims of this paper, 
namely, to derive the Dirac equation with chiral symmetry and discuss its far reaching 
physical implications.   Throughout this paper, we refer to our new equation as the 
Dirac equation with chiral symmetry (DECS) to make distinction from the commonly 
used name 'generalized Dirac equation' in many published papers.

Different generalizations of the Dirac equation (DE) were previously presented and, 
in general, they can be divided into two cathegories: those that derived the equation
and those that just add {\it ad hoc} terms to it.  More specifically, some obtained 
equations were used to either unify leptons and quarks [11-13], or account for 
different masses of three generations of elementary particles [14-16], or extend the 
mass term to include {\it ad hoc} a pseudoscalar mass [17] .  In other generalizations, 
the DE was derived for distances comparable to the Planck length [18], or with the 
external magnetic field included [19], or even for higher integer and half-integer 
spins [20], which requires combining the Dirac [8] and Klein-Gordon [21,22] 
equations.  Another generalization of the DE involved changing a phase factor in 
four-component spinors, which allowed for different masses in the equation [23].  

The Dirac equation with chiral symmetry derived in this paper is new and its physical 
implications are far reaching.  First, our derivation of the DECS demonstrates that this
equation can be obtained from the eigenvalue equation that represents the condition 
required by the four-component spinors to transform as one of the irreps of the 
Poincar\'e group $\mathcal {P}$ extended by parity.  Second, the eigenvalue equation 
allows deriving either the DE or the DECS.  Third, the DE is obtained by factorization 
of the Klein-Gordon equation if, and only if, a specific choice of chirial basis is selected.  
Fourth, as compared to the DE, there is an extra mass term in the DECS and its 
properties allow identifying it with pseudo-scalar mass.  

Our formal derivation of the pseudo-scalar mass term and relating it directly to 
chirial symmetry gives physical justification for the existence of this term and 
allows us to discuss its physical implications; this makes our approach so 
different from the previuosly {\it ad hoc} addition of pseudo-scalar mass to
the DE without neither justifying its physical presence nor its origin [17].  
Our obtained results demonstrate that this pseudo-scalar mass in the DECS, 
its relationship to chiral symmetry and the resulting pseudo-scalar Higgs can 
be used to explain smallness of neutrino masses [24,25], and also properties
Dark Matter (DM) particles [26,27].

The theory presented in this paper emphasizes the distinction between 
the Lorentz-invariant (but non-conserved) concept of chirality and the 
non-Lorentz-invariant (but conserved) property of helicity for massive 
neutrinos. The neutrino considered here is a left-handed Dirac particle. 

The paper is organized as follows: the eigenvalue equation for bispinors are given
in Section2; the fundamental equation for bispinors with chiral symmetry is derived
in Section 3;  physical implications of the obtained results are discussed in Section 4;
and conclusions are presented in Section 5.

\section{Eigenvalue equation for bispinors}

The condition that a scalar wavefunction $\phi$ transforms as one of the irreps 
of $T(3+1) \subset \mathcal {P}$ is given by the following eigenvalue equation 
[6,7]
\begin{equation}
i \partial_{\mu} \phi \ =\ k_{\mu} \phi\ ,
\label{eq1}
\end{equation}
\noindent
where $k_{\mu}$ labels the irreps.  To generalize this result to Dirac spinors,
called also bispinors, we follow Wigner [1,2], who proved that the proper irreps 
of spin-1/2 elementary particles are the four-component bispinors $\psi$ for 
which the eigenvalue equation given by Eq. (\ref{eq1}) becomes 
\begin{equation}
i A^{\mu} \partial_{\mu} \psi \ =\ A^{\mu} k_{\mu} \psi\ ,
\label{eq2}
\end{equation}
\noindent
where $A^{\mu}$ is an arbitrary constant matrix of $4 \times 4$.  Defining 
$X^{\mu} = -i A^{\mu}$ and $Y = A^{\mu} k_{\mu}$, we obtain 
\begin{equation}
\label{eq3}
(X^{\mu} \partial_{\mu}+Y) \psi = 0\ ,
\end{equation}
with $X^{\mu}$ and $Y$ to be determined.   This is the general condition 
for the bispinors to transform as one of the irreps of $\mathcal {P}$ and 
this condition will be now used to derive the DECSfor the four-componet 
bispinors given by   
\begin{equation}
\psi =  \begin{bmatrix}
    \chi_{L} \\
    \chi_{R}\\
\end{bmatrix}
\label{eq4}
\end{equation}
where $\chi_{L}$ and $\chi_{R}$ are two component of bispinors.  The 
necessity of coupling these spinors is understood mathematically as 
accommodating the sign ambiguity introduced in the construction of the 
isomorphism between boosts in SO(3,1) and those in SU(2).   As a result 
of this, we find $\chi_{L}$ and $\chi_{R}$ to transform identically under 
rotations, but oppositely under boosts [3].  We may thus write our Lorentz 
transformation for the bispinors as
\begin{equation}
\Lambda = { \begin{bmatrix}
  \Lambda_{L}   &   0  \\
  0   &   \Lambda_{R}  \\
  \end{bmatrix}} = \begin{bmatrix}
  \exp{\big(\frac{i\vec{\sigma} \cdot (\vec{\theta} - i \vec{\phi})}{2}\big)}   &   0  \\
  0   &   \exp{\big(\frac{i\vec{\sigma} \cdot (\vec{\theta} + i \vec{\phi})}{2}\big)}  \\
  \end{bmatrix}
\label{eq5}
\end{equation}
where $\vec{\theta}$ and $\vec{\phi}$ parameterize our rotations and boosts, respectively, 
and are related to the transformations of four vectors via the four-by-generators $\vec{J}$ 
and $\vec{K}$ such that
\begin{align*}
\hat{\Lambda}=\exp{\bigg( i \vec{J} \cdot \vec{\theta} + i \vec{K}\cdot\vec{\phi}\bigg)}\ .
\label{eq6}
\end{align*}

\section{Fundamental equation for bispinors with chirial symmetry}

To determine forms of the matrices $X^{\mu}$ and $Y$, the Lorentz transformation must 
be applied to the RHS and LHS of Eqs. (\ref{eq4}) to (\ref{eq3}), and this yields
\begin{equation} 
\big((\Lambda^{-1} \hat{\Lambda}^{\nu}_{\mu} X^{\mu} \Lambda)\partial_{\nu}+
(\Lambda^{-1} Y \Lambda) \big)\psi = 0\ . 
\label{eq7}
\end{equation}
This leads to the necessary conditions for invariance
\begin{equation}
\hat{\Lambda}^{\nu}_{\mu} X^{\mu} = \Lambda X^{\nu} \Lambda^{-1} \qquad  
\hskip0.25in {\rm and} \hskip0.25in  Y = \Lambda Y \Lambda^{-1}\ .
\label{eq8}
\end{equation}
Solving these, we find the most general form of our matrix coefficients written in block form
\begin{equation}
    X^\mu = {\begin{bmatrix}
        0   &   x_{R} (\sigma^{0} \delta^{\mu}_{0} + \sigma^{k} \delta^{\mu}_{k})  \\
        x_{L} (\sigma^{0} \delta^{\mu}_{0} - \sigma^{k} \delta^{\mu}_{k})   &   0  \\
\label{eq9}
    \end{bmatrix}}
\end{equation}
\begin{equation}
    Y = {\begin{bmatrix}
        y_{L}\sigma^{0}   &   0                \\
                     0    &   y_{R}\sigma^{0}  \\
\label{eq10}
    \end{bmatrix}} 
\end{equation}
where $x_{R}$, $x_{L}$, $y_{R}$, and $y_{L}$ are free parameters.  

Taking the Dirac $\gamma$ matrices in the Weyl basis
\begin{equation}
    \gamma^{0} = {\begin{bmatrix}
        0    &   \sigma^{0}                \\
        \sigma^{0}    &   0  \\
    \end{bmatrix}} 
    \qquad
    \gamma^{k} = {\begin{bmatrix}
        0    &   \sigma^{k}                \\
        -\sigma^{k}    &   0               \\
    \end{bmatrix}}
\end{equation}
and identifying the chiral projection operators as
\begin{equation}
    P_{L} = {\begin{bmatrix}
        \sigma^{0}     &   0                \\
        0    &   0  \\
    \end{bmatrix}} 
    \qquad
    P_{R} = {\begin{bmatrix}
        0     &   0                \\
        0    &   \sigma^{0}  \\
    \end{bmatrix}} 
\end{equation}
we obtain
\begin{equation}
\big((x_{L} P_{R} + x_{R} P_{L})\gamma^{\mu} \partial_{\mu} + (y_{L}P_{L} 
+ y_{R}P_{R})\big)\psi = 0\ .
\label{eq11}
\end{equation}
Now, under the assumption that $x_{L}$ and $x_{R}$ are nonzero we are free to 
multiply from left with $i(x_{L} P_{R} + x_{R} P_{L})^{-1}$, and obtain
\begin{equation}
\bigg(i\gamma^{\mu} \partial_{\mu} + i\bigg(\frac{y_{L}}{x_R}P_{L} + \frac{y_{R}}
{x_L}P_{R}\bigg)\bigg)\psi = 0\ ,
\label{eq12}
\end{equation}
which shows that there are only two independent degrees of freedom in the derived 
equation.  To identify the physical basis of these degrees, we observe that Eq. (\ref{eq12}) 
gives the following squared-Hamiltonian
\begin{equation}
\mathcal{H}^2 \psi = \bigg(\partial^{k} \partial_{k} - \frac{y_{L} y_{R}}{x_{L} x_{R}}
\bigg)\psi\ ,
\label{eq13}
\end{equation}
and thus we find the emergence of the propagation mass term
\begin{equation}
m \equiv \pm i \sqrt{\frac{y_{L} y_{R}}{x_{L} x_{R}}}\ .
\label{eq14}
\end{equation}
The restriction of the square of Eq. (\ref{eq14}) to positive real numbers is equivalent 
to the physical restriction of Einstein energy-momentum relationship.  Our remaining 
degree of freedom may be identified with the choice of a chiral basis.  Let us define 
the chiral angle as
\begin{equation}
\begin{aligned}
 \alpha \equiv -\frac{i}{2} \ln{\bigg(\mp  \sqrt{\frac{x_L y_L} {x_R y_R}}}\bigg)\ .
\end{aligned}
\label{eq15}
\end{equation}
Thus, the final compact form of our Dirac equation with chirial symmetry is
\begin{equation}
\begin{aligned}
(i\gamma^\mu \partial_\mu - m e^{-2i \alpha \gamma^5})\psi = 0\ . 
\end{aligned}
\label{eq16}
\end{equation}
Since this equation is Poincar\'e invariant, it is the fundamental equation 
of physics.  This equation reduces to the original Dirac equation when 
$\alpha = 0$.

\section{Chiral symmetry and pseudoscalar mass}

A consequence of Eq. (\ref{eq16}) is made explicit by proffering an alternative, 
suggestive parameterization.  Let us define 
\begin{equation}
\begin{aligned}
M\equiv -\frac{i}{2}\bigg( \frac{y_R}{x_L} + \frac{y_L}{x_R}\bigg)= m \cos{2\alpha}\ , 
\hspace{7mm} 
\widetilde{M} \equiv -\frac{i}{2}\bigg( \frac{y_R}{x_L} - \frac{y_L}{x_R}\bigg) = -im 
\sin{2\alpha}\ ,\\
\end{aligned}
\label{eq17}
\end{equation}
so that Eq. (\ref{eq16}) becomes
\begin{equation}
\big(i \gamma^{\mu} \partial_{\mu} - M - \widetilde{M} \gamma^{5} \big)\psi = 0\ .
\label{eq18}
\end{equation}
This form admits the simultaneous validity of both fundamental scalar and fundamental 
pseudoscalar mass terms.  To better understand these mass terms, consider a global 
chiral transformation given by
\begin{equation}
\psi \rightarrow \psi' = e^{i \gamma^{5} \beta} \psi\ .
\label{eq19}
\end{equation}
The Lagrangian of Eq. (\ref{eq19}) may be written as
\begin{equation}
\mathcal{L} = i \bar{\psi} \gamma^{\mu} \partial_{\mu} \psi - \bar{\psi}(M+
\widetilde{M}\gamma ^5) \psi\ , 
\label{eq20}
\end{equation}
and under the transformation of (\ref{eq19}) this becomes
\begin{equation}
\mathcal{L'} = i \bar{\psi} \gamma^{\mu} \partial_{\mu} \psi - \bar{\psi}(M+
\widetilde{M}\gamma ^5) e^{-2i\beta \gamma^5} \psi\ ,
\label{eq21}
\end{equation} 
so we the effect of (\ref{eq19}) is equivalent to the rotation of our mass 
parameters
\begin{equation}
{\begin{bmatrix}
        M'   \\
        i\widetilde{M}' \\
    \end{bmatrix}}
    =
    {\begin{bmatrix}
        \cos{2 \beta}    &   -\sin{2 \beta}                \\
        \sin{2 \beta}   &   \cos{2 \beta}               \\
    \end{bmatrix}}
{\begin{bmatrix}
        M   \\
        i\widetilde{M} \\
    \end{bmatrix}}
    =
    {\begin{bmatrix}
        m \cos[{2(\alpha +\beta)}]  \\
        m \sin[{2(\alpha +\beta)}] \\
    \end{bmatrix}}
\label{eq22}
\end{equation}
This is precisely the transformation required to leave invariant the propagation mass in 
Eq. (\ref{eq16}).  Note that for a single field, we may always apply a chiral rotation so 
as to reorient our chiral axes and force the pseudoscalar mass to vanish.   However, this 
is not a valid procedure in the case of multipartite systems whose constituent fields obey 
Eq. (\ref{eq16}) for different chiral anlges.  Therefore, we conclude that it is only the 
nonequivalence of chiral basis that gives rise to physically observable effects. 

We found that the chiral rotation of a massive field is equivalent to an alternative choice 
of chiral basis.  This, in turn, is an alternative factorization of the Klein-Gordon equation 
(i.e. the Einstein energy relationship).  These nonstandard factorizations necessarily 
redistribute the fraction of the mass available to the field between the  left- and right-chiral 
components.  Our work presented in this paper significantly differs from the previous studies
of both chirial symmtery and generalization of Dirac equation.  

Our main result is first-principle proof of the necessity of specifying the chiral basis in which 
the fundamental induced representations of the Poincar\'{e} group reside.  This is not a mere 
mathematical triviality, but rather the necessary consequence of considering fundamental 
physical symmetries in flat space-time and its anticipated physical implications that are now
presented and discussed.

\section{Physical implications}

\subsection{Yukawa-generated masses}

As demonstrated above, together $m$ and $\alpha$ define the infinite parameter space in 
which all first-order Poincare-invariant equations for massive spin-1/2 particles must reside.  
The simplest and most physically-motivated selection process for the determination of $m$ 
and $\alpha$ is obtained by the interpretation of scalar and pseudoscalar mass terms as 
having their origin in Yukawa couplings toreal scalar fields with non-vanishing vacuum 
expectation values (VEVs). 

Treating these fields independently and expanding them about the ground state, we 
may write the relevant contributions of these coupling to the Lagrangian as
\begin{equation}
\begin{aligned}
\mathcal{L}_{Y} &= -\lambda_1 \bar{\psi} \phi_1 \psi - \lambda_2 \bar{\psi} \phi_2 
\gamma^{5}\psi \\&\approx - \frac{\lambda_1 v_1}{\sqrt{2}} \bar{\psi}\psi- 
\frac{\lambda_2 v_2}{\sqrt{2}} \bar{\psi}\gamma^{5}\psi\ ,
\label{eq23}
\end{aligned}
\end{equation}
where the $v_1$ and $v_2$ are the VEVs of the two fields $\phi_1$ and $\phi_2$, which 
are treated here as independent.  It is easy to restrict this model to encompass a single field 
with a complicated set of coupling parameters.  The coupling constants $\lambda_1$, 
$\lambda_2$ and VEVs are related to our parameters $m$ and $\alpha$ via
\begin{equation}
\begin{aligned}
m &= \sqrt{\frac{\lambda_1^2 v_1^2 - \lambda_2^2 v_2^2}{2}} \\
\alpha  &= \frac{i}{4}\ln{\bigg(\frac{\lambda_1 v_1 + \lambda_2 v_2}{\lambda_1 
v_1 - \lambda_2 v_2}\bigg)}\ ,\\
\label{eq24}
\end{aligned}
\end{equation}
which allows both $\alpha$ and $m$ to be completely fixed by the VEVs of the fields 
$\phi_1$ and $\phi_2$ and their respective coupling constants.

\subsection{Neutrino masses}

The presented theory highlights the chiral angle as the necessary consequence of the 
identification of particle states with the induced irreducible representations of the Poinca.\'ere 
group.  This degree of freedom is necessarily present in all fermions of the Standard Model. 
However, the non-observation of right-handed neutrinos (left-handed antineutrinos) in 
Nature permits motivates applications of the theory to neutrinos, which are now described.

In the context of Eq. (\ref{eq23}), the case of neutrino fields takes on an interesting dimension 
by offering a potential solution to an open problem.   In the Standard Model [3], right-chiral 
leptons form singlets that preclude the existence of right-chiral neutrinos.  However, if this 
condition is enforced, then the neutrino field $\nu$ becomes
\begin{equation}
\frac{1}{2}(1-\gamma^5)\nu = \nu  \hspace{5mm}
\frac{1}{2}(1+\gamma^5)\nu = 0\ .
\label{eq25}
\end{equation}
Combining these constraints with Eq. (\ref{eq23}), we find 
\begin{equation}
\begin{aligned}
\mathcal{L}_Y &= -\frac{1}{2\sqrt{2}} \bar{\nu} (\lambda_1 v_1 -\lambda_2 v_2)\nu\ .
\end{aligned}
\label{eq26}
\end{equation}
Thus, the neutrino mass in the Dirac Lagrangian appears as
\begin{equation}
\begin{aligned}
m_{\nu} &= \frac{1}{2\sqrt{2}} (\lambda_1 v_1 -\lambda_2 v_2)\ .
\end{aligned}
\label{eq27}
\end{equation}

The obtained results demonstrate that the mass of a left-chiral field generated by coupling 
to the Standard Model (SM) Higgs field may be suppressed by a non-zero real pseudoscalar 
coupling to an external scalar field with non-vanishing VEV.   This is clearly of potential 
relevance when considering why the observed masses of neutrinos vary from those of the 
other fermions by orders of magnitudes.   Many different explanations have been offered 
to account for this discrepancy [24,25], but the mechanism presented here is particularly 
appealing as it suggests the smallness of the neutrino masses is the necessary consequence 
of the non-existence of right-chiral neutrinos; more detailed calculations of this phenomenon
will be presented elsewhere.

Let us point out that the most immediate testable consequence of additional neutrino-scalar 
couplings is the necessary modification of elastic scattering cross sections.   While in the 
absence of interference phenomena it is likely for modifications to the SM predictions to be 
sub-leading, evidences of deviation from SM predictions have already been obtained by the 
LSND [28] and MiniBooNE [29] collaborations.  So far these anomalies have refuted explanation.  
Howevre, additional precision measurements being preformed by the MicroBooNE collaboration 
and to-be-performed by the DUNE collaboration will provide further constraints and may--given 
long enough livetimes--obtain constraints sufficient to definitively probe the validity of the model 
proposed. 

\subsection{Dark matter candidate}

Another attractive aspect of this mass-generating mechanism is the natural inclusion of a dark 
matter candidate.  The evidence for the existence of particle dark matter is strongly supported 
by the astronomical observations of galactic rotational curves and stability of clusters of galaxies 
[31,32] and while there have been numerous theories postulated different elementary particles 
[27,33,34], so far the existence of those particles have not yet been verified experimentally 
[35-37].   As possible candidates, weakly interacting massive particles (WIMPs), supersymmetric 
(SUSY) partciles like neutralinos [33], axions [38], and extermely light bosonic particles (ELBPs) 
suggested by [39,40] and showed to be physically unacceptable by [41,42].

All evidence suggests dark matter couples to ordinary matter primarily through mass-mass terms, 
ostensibly gravitational [43,44].  All other couplings are, at most, sub-leading contributions to DM 
interactions.  Thus, the existence of a chiral mass may offer insight into these observations in several 
ways.  First, it is possible that pseudoscalar mass contribution to the gravitational field differs from 
those of scalar mass terms; an in-depth investigation of that possibility is beyond the scope of this 
paper.   Second, the results obtained in this work make also plausible the existence of a massive 
$\phi_2$ particle that couples minimally and exclusively to chirally-asymmetric fields (neutrinos) 
and thereby satisfies many of the required characteristics of dark matter: a minimal non-gravitational 
coupling with a restricted phase space for decay resulting in a long-lived particle [45].  The contribution 
of this postulated field to gravitational interactions depends on the mass and abundance of the postulated 
particle, both of which cannot be determined by the presented theory. 

As a last remark, we note a distinguishing characteristic of this model is the wide range of allowable 
inertial (and thus gravitational) masses consistent with Higgs-portal and direct-detection observations.  
For massive DM particles, we must amend the aforementioned exclusivity of interactions to include 
scalar as well as chiral-asymmetric couplings.  Then, we must allow the mass of the postulated 
particle to be generated by couplings to both $\phi_1$ (nominally the standard model Higgs) 
and itself. This method of mass generation allows for enhancement of the inertial mass and 
admits a mass parameter that differs non-trivially from its Higgs-coupling parameter.  Thus, the 
proposed model may plausibly span the intermediate parameter space inaccessible to both ELBP 
and WIMP models.  Intriguingly, a non-zero reciprocal coupling between $\phi_1$ and $\phi_2$ 
may have important implications for the renormalization of the Higgs self-energy in the absence 
of supersymmetric models.

\section{Conclusions}

In this paper, a group-theoretical derivation of the most general Poincaré invariant Dirac 
equation with chirial symmetry, which introduces pseudo-scalar mass to the equation, is 
presented.   The derivation is based on the eigenvalue equation that represents the condition 
required by the four-component spinors to transform as one of the irreps of the Poincar\'e 
group.   It is shown that the chiral rotation of a massive field is equivalent to a choice of 
chiral basis and that this choice must be specified in order to factorize the Klein-Gordon 
equation and derive the original Dirac equation.  

The derived Dirac equation with chirial symmetry has two additional degrees of freedom, 
which are identified with the propagation mass ($m$) and the chiral angle ($\alpha$).  
Being a physical parameter, $\alpha$ specifies the chiral basis of the spinor solutions 
and therefore its value must be physically justified.  The existence of the pseudo-scalar 
mass generating Higgs-like couplings is an interesting phenomenon as this pseudo-scalar 
Higgs when combined with the suppression of right-chiral neutrinos offers a natural 
mechanism for producing anomalously small left-chiral neutrino masses.  Moreover,
this psuedo-scalar Higgs field satisfies many of the requirements for a DM candidate,
whose existence can be verified experimentally. 

The derived Dirac equation with chiral symmetry applies to all fermions in the 
Standard Model because the introduced chiral angle by the equation is present in all 
particles with spin $1/2$.  Therefore, once interations are included into the theory, its 
further integration with the Standard model will be done and reported separately.

\section*{Acknowledgements}
We are grateful to an anonymous referee for providing comments and suggestions 
that allowed us to improve the revised version of this paper.  We thank B. Jones for 
his comments on our first draft of this paper and for bringing to our attention the 
paper by Leiter and Szamosi (1972).  ZEM also thanks for a partial support of this 
research by the Alexander von Humboldt Foundation. 


\end{document}